# Compositional Correlation for Detecting Real Associations Among Time Series

## Fatih DIKBAS

Civil Engineering Department, Pamukkale University, Turkey

Correlation remains to be one of the most widely used statistical tools for assessing the strength of relationships between data series. This paper presents a novel compositional correlation method for detecting linear and nonlinear relationships by considering the averages of all parts of all possible compositions of the data series instead of considering the averages of the whole series. The approach enables cumulative contribution of all local associations to the resulting correlation value. The method is applied on two different datasets: a set of four simple nonlinear polynomial functions and the expression time series data of 4381 budding yeast (saccharomyces cerevisiae) genes. The obtained results show that the introduced compositional correlation method is capable of determining real direct and inverse linear, nonlinear and monotonic relationships. Comparisons with Pearson's correlation, Spearman's correlation, distance correlation and the simulated annealing genetic algorithm maximal information coefficient (SGMIC) have shown that the presented method is capable of detecting important associations which were not detected by the compared methods.

**Keywords:** Saccharomyces Cerevisiae, Gene Expression, Bioinformatics, Combinatorics, Compositional Variance, Compositional Covariance, Compositional Correlation

## 1. INTRODUCTION

Pearson's correlation coefficient (Pearson 1895) (simply called correlation) might still be the most widely used statistical measure for assessing relationships between data series. Nevertheless, there are many warnings in literature about its improper use: Correlation is misleading (Bland and Altman 1986); good correlation does not automatically imply good agreement (Lobbes and Nelemans 2013); risk of producing spurious correlations when analyzing non-independent variables is very large (Brett 2004, Duan et al. 2014).

Four problems need to be addressed if the Pearson's correlation coefficient is used to characterize the dependence between two time series (Pasanen and Holmström 2016):

— if both time series contain a temporal trend, they might be deemed correlated even though they are not related in any way;

— correlation might change in time, in which case the Pearson's correlation coefficient only reflects average correlation over the length of the time series;

— such a time window-related concept of scale causes a situation where the data sets contain different correlation structures when viewed in different time horizons;

— difficulty of making valid statistical inference about the correlation.

Despite all these warnings, Pearson's correlation is also used in the analysis of high-throughput data (such as genotype, genomic, imaging, and others) (Langfelder and Horvath 2012, Rachel Wang et al. 2015) although the relationships are generally nonlinear. The occasionally misleading and insufficient results of Pearson's correlation cause the subsequent investigations like cluster analysis to become partially obsolete, deficient and unreliable as shown with many examples presented in this paper. The unintended and generally unnoticed misleading results of Pearson's correlation are caused by the approach used in calculation of the correlation itself where the averages of the whole series are used for assessing relationship. In fact, the average value of a data series is a single value which does not reflect the variations within the data series while the variations might have great importance in the determination of associations which generally vary through the time course of the compared data series. The idea behind the presented compositional correlation method is that the global dependence structure is better described by a combination of all local measures of dependence computed in different regions composing the whole series than a single-number measure of dependence. The method is based on the foundations of the two-dimensional correlation method developed by the author for assessing the degree and direction of relationships between matrices (Dikbas 2017b, Dikbas 2018). The two-dimensional correlation method calculates the horizontal correlation by considering the averages of all rows and the vertical correlation by considering the averages of all columns in the compared matrices instead of the overall averages of the matrices. The need for developing these methods were realized when it was noticed that, in some cases, the Pearson's correlation value decreased even though the estimations became closer to the observations in the hydrological estimation studies (Dikbas 2016a, Dikbas 2016b, Dikbas 2017a). It was understood that the problems with the Pearson's correlation was caused by taking the average of the whole series. With a similar approach used in the two-dimensional correlation method, the calculation of compositional correlation considers the averages of all parts of all possible compositions of the data series instead of considering the averages of the whole series. The process is based on compositional variance which reflects the variation of the spread of the values in the series from the local means. The approach enables determination of variation dependent compositional correlations between two data series and consequently provides valuable information for assessing new relationships some of which are impossible to be detected when widely used correlation approaches are applied.

## 2. METHODS

### 2.1 Influence of sample size on Pearson's correlation

One of the most important factors that influence the value of Pearson's correlation coefficient (r) is the size (n) of the sample data. Both of the compared data series should have at least two elements to be able to calculate correlation ($n \geq 2$). There are six types of possible relationships when both of the series have two elements: When n = 2, the correlation value can either be 1 (when both series are increasing or decreasing together), -1 (when one series is increasing and the other series is decreasing) or undefined (when one series is increasing or decreasing and the other is constant or both are constant) regardless of the values in the series (Figure 1). All data series pairs for which a correlation can be calculated are a combination of the six relationship types in Figure 1. The value of correlation is directly related with the composition structure of the six types in the figure. Pearson's correlation perfectly determines the relationships when n = 2 but the problems associated with it start when n > 2.

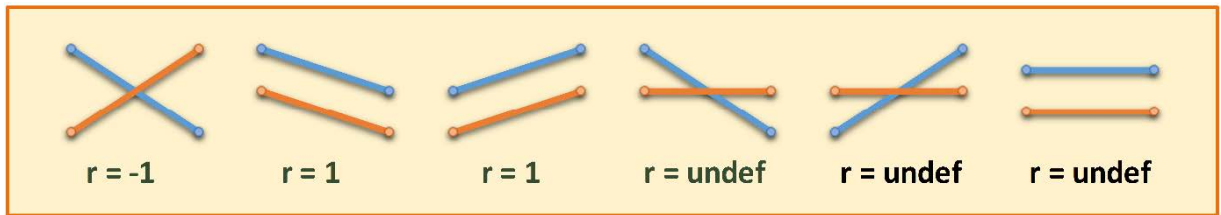

**Figure 1** Correlations between all possible relationships between two data series having two values.

An example showing the variation of Pearson's correlation according to the selection of observation will be given by using the gene expression example presented in Figure 2. Both series have 23 observations (n=23) and r = -0.12 between the whole time series of the genes. The correlation value decreases from 1 to -0.12 when observation periods from 1-2 to 1-23 are considered (r = 1 for the range 1-2; r = 0.91 for the range 1-3; r = 0.95 for the range 1-4; . . . and r = -0.12 for the range 1-23 as shown with the blue line in Figure 2b). The correlation between the first five observations is 0.99 and r suddenly decreases to 0.22 when the first six observations are considered. Value of r continues to have very low values for the remaining ranges and does not get positive values for the ranges from (1-7) to (1-23).

When the first five observations are ignored (for which r = 0.99) the r values change between 1 (for the range 6-7) and 0.48 (for the range 6-23) Figure 2b). The figure clearly shows that correlation only gets negative values when the first five values which nearly have a perfect positive correlation are included in the calculation of correlation. The expression time series of the sample genes always increase and decrease together for all smallest subsections (n = 2) (Figure 2a) indicating a very strong numerical relationship but the Pearson's correlation does not reflect this behavior.

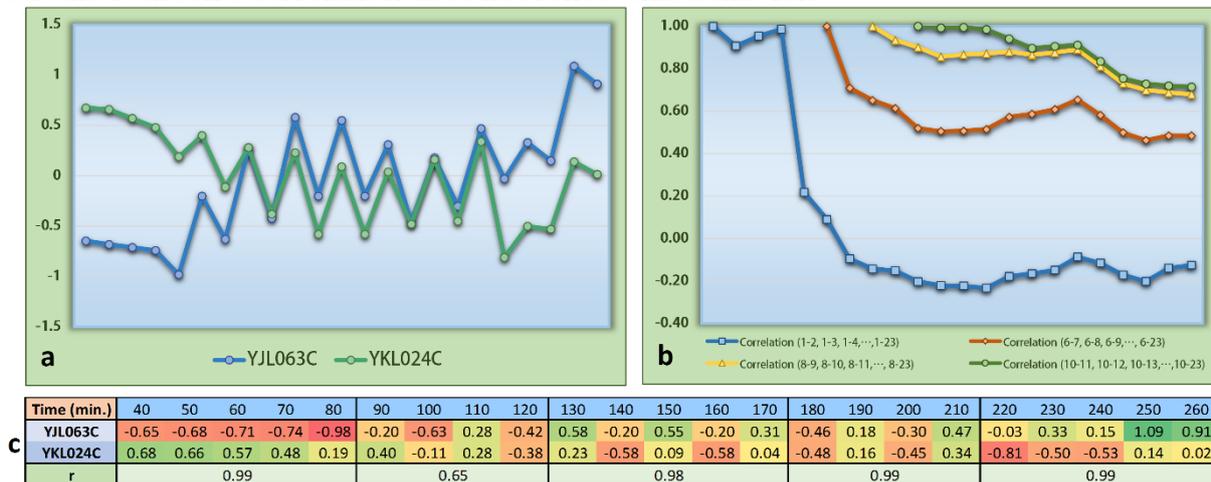

**Figure 2 (a)** The expression time series of budding yeast genes YJL063C and YKL024C, **(b)** the variation of Pearson's correlation with the selected data range and **(c)** the expression series and the correlations between the parts of the BCC of the genes YJL063C and YKL024C.

The developed compositional correlation method calculates correlations for all compositions of the compared series and the values tend to be higher when the parts of the compositions are highly correlated. For example, the above gene pair is one of the numerous gene pairs determined to have a very low value of Pearson's correlation while most of the compositional correlations are very high (up to 0.92). For this pair, the best correlated composition (BCC) is [5, 4, 5, 4, 5]. When the 23 observations are divided into 5 parts (which form the BCC) as shown in Figure 2c, r = 0.65 for the second part and r ≥ 0.98 for the remaining parts. The high r values in all parts point out a strong direct relationship between the series as depicted by the time series graph (Figure 2a) while the r value for the whole series is interestingly a negative number close to zero indicating an inverse weak relationship. The high value of the compositional correlation for the gene pair indicates the apparent direct relationship. The above example was only one of the numerous gene pairs detected in this study showing the influence of sample size on the value of correlation.

The compositional correlation approach enables detection of this type of relationship by considering the cumulative influence of all possible sample compositions for the compared series. If there is no composition producing high correlations for the compared parts, then the compositional correlation will not be high showing that the association between the compared series is definitely weak.

## 2.2 Calculation of compositional correlation

A composition of an integer n is a vector in which the components are positive integers that sum to n. Each component of a composition is called a part of the composition. For instance, the compositions of 3 are [1, 1, 1], [1, 2], [2, 1] and [3]. Similarly, if a sample data series has n (a positive integer) elements, we can determine the compositions of the series by dividing it into parts. Calculation of compositional correlation requires that each part should have at least 2 elements (m ≥ 2). Table 1 shows the compositions for 2 ≤ n ≤ 10 with parts ≥ 2. The numbers in the brackets are the number of elements in each part and the column on the right shows the total number of compositions, $t_n$.

**Table 1** All possible compositions with parts ≥ 2 for 2 ≤ n ≤ 10

| n | All possible compositions with parts ≥ 2 | $t_n$ |
|---|---|---|
| 2 | [2] | 1 |
| 3 | [3] | 1 |
| 4 | [2, 2]; [4] | 2 |
| 5 | [2, 3]; [3, 2]; [5] | 3 |
| 6 | [2, 2, 2]; [2, 4]; [3, 3]; [4, 2]; [6] | 5 |
| 7 | [2, 2, 3]; [2, 3, 2]; [2, 5]; [3, 2, 2]; [3, 4]; [4, 3]; [5, 2]; [7] | 8 |
| 8 | [2, 2, 2, 2]; [2, 2, 4]; [2, 3, 3]; [2, 4, 2]; [2, 6];<br>[3, 2, 3]; [3, 3, 2]; [3, 5]; [4, 2, 2]; [4, 4]; [5, 3]; [6, 2]; [8] | 13 |
| 9 | [2, 2, 2, 3]; [2, 2, 3, 2]; [2, 2, 5]; [2, 3, 2, 2]; [2, 3, 4]; [2, 4, 3];<br>[2, 5, 2]; [2, 7]; [3, 2, 2, 2]; [3, 2, 4]; [3, 3, 3]; [3, 4, 2]; [3, 6];<br>[4, 2, 3]; [4, 3, 2]; [4, 5]; [5, 2, 2]; [5, 4]; [6, 3]; [7, 2]; [9] | 21 |
| 10 | [2, 2, 2, 2, 2]; [2, 2, 2, 4]; [2, 2, 3, 3]; [2, 2, 4, 2]; [2, 2, 6]; [2, 3, 2, 3]; [2, 3, 3, 2]; [2, 3, 5];<br>[2, 4, 2, 2]; [2, 4, 4]; [2, 5, 3]; [2, 6, 2]; [2, 8];<br>[3, 2, 2, 3]; [3, 2, 3, 2]; [3, 3, 2, 2]; [3, 2, 5]; [3, 3, 4]; [3, 4, 3]; [3, 5, 2]; [3, 7];<br>[4, 2, 2, 2]; [4, 2, 4]; [4, 3, 3]; [4, 4, 2]; [4, 6];<br>[5, 2, 3]; [5, 3, 2]; [5, 5]; [6, 2, 2]; [6, 4]; [7, 3]; [8, 2]; [10] | 34 |

The number of possible compositions rapidly increases with n. The total numbers of compositions constitute a Fibonacci sequence ($t_n = F_{n-1}$). An interesting feature of Fibonacci numbers which are defined by $F_{n+1} = F_n + F_{n-1}$ is that the rate $F_{n+1} / F_n$ rapidly tends to the golden ratio known as $\varphi = (1 + \sqrt{5})/2 = 1.618...$ (Stakhov 1989). This means that there is golden ratio between the total number of compositions for two consecutive integers when the minimum number of observations in each part is equal to 2 (m = 2).

### 2.2.1. Compositional Variance.
Before presenting the relationship for compositional correlation, compositional variance should be defined as it constitutes the base of compositional correlation.

Variance is the measure of how far the numbers in a set are spread from their average. The compositional correlation theory presented in this paper is based on the idea that, when the averages of the parts of the compositions of a data series are considered, a variable might have different variances even though the variable has a single variance value when the average of the whole series is considered. In this view, the compositional variance of a time series data is a measure of how far the averages of the parts of a composition are spread from the overall average. Consequently, the compositional variance becomes a cumulative measure of how far the numbers in the time series are spread from their part averages. The following equation defines the compositional variance of a scalar time series for any composition with k parts:

$$Var_c(A) = \frac{\sum_{i=1}^{k} \sum_{j=1}^{n_i} \left(A_{i,j} - \overline{A_i}\right)^2}{n} \tag{1}$$

In this equation:
A: A scalar vector;
n: The number of data in A;
$Var_c(A)$: The compositional variance of the vector [A] for the current composition;
k: The number of parts in the current composition for which the correlation is being calculated;
$n_i$: The number of data in part i;
$A_{i,j}$: The $j_{th}$ data in $i_{th}$ part of vector A;
$\overline{A_i}$: The arithmetic mean of the $i_{th}$ part of vector A;

A compositional variance value of zero indicates that each part of the composition has its own identical value. All variances that are non-zero will be positive numbers. A large compositional variance indicates that the part averages in the set are far from the overall average and from each other, while a small compositional variance indicates the opposite.

Accordingly, the compositional standard deviation may be defined as the square root of the compositional variance. The compositional standard deviation has the same unit with the evaluated data. Because of this property, one might prefer compositional standard deviation for evaluating dispersions from the average.

**2.2.2. Compositional Covariance.** Covariance is a linear gauge of dependence between variables and provides a measure of the strength of the co-variation between variables. Compositional covariance is a measure of how changes in the part averages of a time series are associated with changes in the part averages of a second time series. If the relationship between the part averages is inverse then the compositional covariance will be negative and if the relationship is direct then the compositional covariance will be positive. Higher covariance values indicate a stronger association. If two variables are independent, their covariance is 0, but, having a covariance of 0 does not imply the variables are independent. Therefore, if the parts of two time series are independent, then the compositional covariance between the time series will be 0. The following equation defines the compositional covariance between scalar matrices A and B:

$$Cov_c(A,B) = \frac{\sum_{i=1}^{k}\sum_{j=1}^{n_i}\left(A_{i,j} - \overline{A_i}\right)\left(B_{i,j} - \overline{B_i}\right)}{n} \tag{2}$$

where:

$B_{i,j}$: The $j_{th}$ data in $i_{th}$ part of vector B;

$\overline{B_i}$: The arithmetic mean of the $i_{th}$ part of vector B;

Compositional variance is a special case of the compositional covariance when the two time series are identical:

$$Cov_c(A,A) = Var_c(A) \tag{3}$$

**2.2.3. Compositional Correlation.** Covariance is a dimensioned measure and it is scale dependent. This causes increase in the covariance value when a variable is increased in scale. Correlation is a scaled version of covariance that takes values between $-1$ and $1$ and it is dimensionless. A correlation of $\pm 1$ indicates perfect linear association and 0 indicates no linear relationship.

Based on the above definitions of compositional variance and compositional covariance, this paper presents a new correlation approach for assessing relationships between time series data, the compositional correlation:

$$r_c = \frac{Cov_c(A,B)}{\sqrt{Var_c(A)Var_c(B)}} \tag{4}$$

The following equation can be used for directly calculating the compositional correlation:

$$r_c = \frac{\sum_{i=1}^{k}\sum_{j=1}^{n_i}\left(A_{i,j} - \overline{A_i}\right)\left(B_{i,j} - \overline{B_i}\right)}{\sqrt{\left[\sum_{i=1}^{k}\sum_{j=1}^{n_i}\left(A_{i,j} - \overline{A_i}\right)^2\right]\left[\sum_{i=1}^{k}\sum_{j=1}^{n_i}\left(B_{i,j} - \overline{B_i}\right)^2\right]}} \tag{5}$$

For calculating compositional correlation, the averages of each part in each composition are considered for a better consideration of the contribution of local associations through the observed series. This approach is based on the idea that any observation might be more related with the average of its neighbors than it is related with the average of the whole series.

The compositional correlations for each possible composition should be calculated for making a complete association investigation between two time-series. For m = 2, there are 34 compositional correlations when n = 10; 7778742049 correlations when n = 50 and more than $2 * 10^{20}$ correlations when n = 100. Generally, when the number of data in the compared series increases, the complexity of their relationship also increases. For example, when some sections of the series are directly related, some sections might have strong inverse relations and it is impossible to determine these types of relationships by calculating only one correlation coefficient between the whole observations of two variables (Embrechts 2002). A well-known example of this is that the correlation between x and y is zero for $y = x^2$ even though half of the series has a perfect direct relationship while the other half has a perfect inverse relationship. A detailed investigation of the compositional correlations for this function is given below together with the investigations of three other polynomial functions with alternating (from direct to inverse and inverse to direct) relationships through the investigated range.

The compositional correlation method provides the user, the ability to determine direct, indirect, linear and nonlinear relationships between two series by considering the contribution of the relationships between all possible parts of the compared series. The composition that gives the highest and lowest compositional correlations (HCC and LCC) are called the best and worst correlated compositions (BCC and WCC). Together with the remaining compositional correlations, the HCC and LCC values for two data series provide valuable information about the relationships between the series because all compositional correlations (including the Pearson's correlation r) vary between HCC and LCC. The compositional correlation is equal to the Pearson's correlation if the number of parts (k) in the composition is equal to 1 and for all data series with any length, one of the compositional correlations is always equal to the Pearson's correlation (when k = 1). The possible values of the compositional correlations vary between -1 (indicating perfect inverse relationship which does not have to be linear) and 1 (indicating perfect direct relationship which also does not have to be linear).

The existing correlation-based measures are more sensitive to outliers than observations near the average (Legates and Davis, 1997) and this oversensitivity to outliers leads to a bias toward the extreme events (Legates and McCabe Jr, 1999). The presented compositional correlation approach has a very important advantage over the traditional correlation approaches in that it restricts the influence of outliers to the part where they belong and an outlier cannot have influence on the calculation of the averages of other periods. Namely, an outlier only influences the average of its own part when the compositional correlations are being calculated but in the traditional correlation approach, an outlier changes the average of the whole series and generally causes a significant change in the correlation between the compared series. This influence becomes especially more important in the evaluation of shorter series.

## 2.3 Compositional correlations for four polynomial functions

Four simple polynomial functions were selected for assessing the properties of compositional correlation when the relationship between the functions and x are not linear. The selected polynomial functions are as follows:

$$f(x) = x^2 \tag{6}$$

$$f(x) = x^3 - x \tag{7}$$

$$f(x) = x^4 - 10x^2 + 9 \tag{8}$$

$$f(x) = x^3 + x^2 + 2x + 4 \tag{9}$$

The first three functions have both increasing and decreasing sections. The increasing sections have direct relationship with x and the decreasing sections are inversely related with x. This alternating relationship is impossible to be detected by calculating a single correlation value for the selected ranges. The first and the third functions are symmetric across the vertical axis and the second function is symmetric across the origin. The fourth function is a monotonically increasing nonlinear function. Figure 3 shows the curves of the polynomials within the investigated ranges. The points on the curves are obtained by dividing the horizontal ranges into 30 equal pieces.

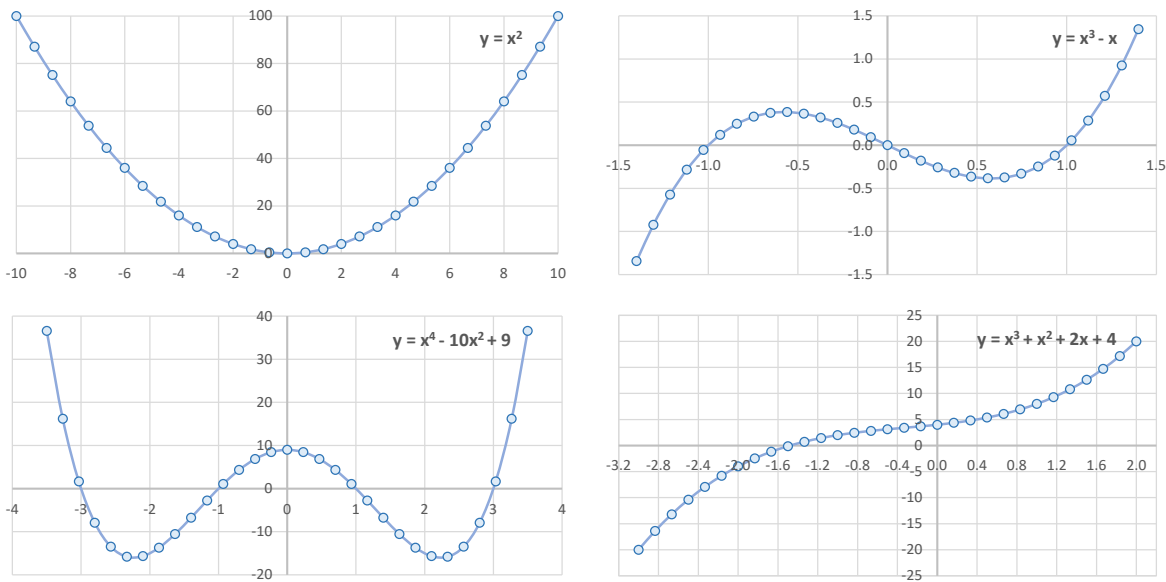

**Figure 3** The curves of the selected four polynomials.

For all polynomials, the investigated ranges are divided into 20 and 30 curve pieces and compositional correlations with x are separately calculated. When the curves are divided into 20 pieces (n = 21), 6765 (the 20th Fibonacci number) compositional correlations are calculated and 832040 (the 30th Fibonacci number) compositional correlations are calculated when the curves are divided into 30 pieces (n = 31). The increasing number of divisions allows a more precise correlation calculation as shown in the obtained results below.

### 2.3.1 Results for y = x²

The presented results are for n = 31. The highest compositional correlation (HCC) 0.9449 is obtained for the composition [2, 2, 2, 2, 2, 2, 2, 2, 15] (the BCC) while the lowest compositional correlation (LCC) -0.9449 is obtained for the composition [15, 2, 2, 2, 2, 2, 2, 2, 2] (the WCC). This is an expected result because the compositional correlation approach exactly determines the directly and inversely related portions of the functions by considering the contribution of the parts of each composition to the overall compositional correlation value. In the case of $y = x^2$, all points on the left of the vertical axis are inversely related with x and therefore they have a negative contribution. Similarly the points on the right of the y axis have a positive contribution to the values of the compositional correlations. Nevertheless, the Pearson's correlation never considers the inversely and directly related portions and produces a zero correlation value for this function which is caused by the equally distributed inversely and directly related curve sections.

Figure 4 summarizes the results obtained for $y = x^2$ when the investigated range is divided into 30 pieces. Each subfigure contains 832040 points. The first two subfigures are the compositional variance clouds for y and x respectively and show the variation of compositional variance with compositional correlation. The uppermost points in the variance clouds represent the values obtained when the average of the whole series is used in the calculation of compositional correlation which is equal to the Pearson's correlation (k = 1, m = 2). As the figures indicate, the compositional variance is maximum when the average of the whole series is used because the total distance between the curve and the horizontal average line becomes maximum. When the curve is divided into subgroups then the compositional variances are lower. The variance clouds are symmetric because half of the curve is inversely related with x and the other half is directly related.

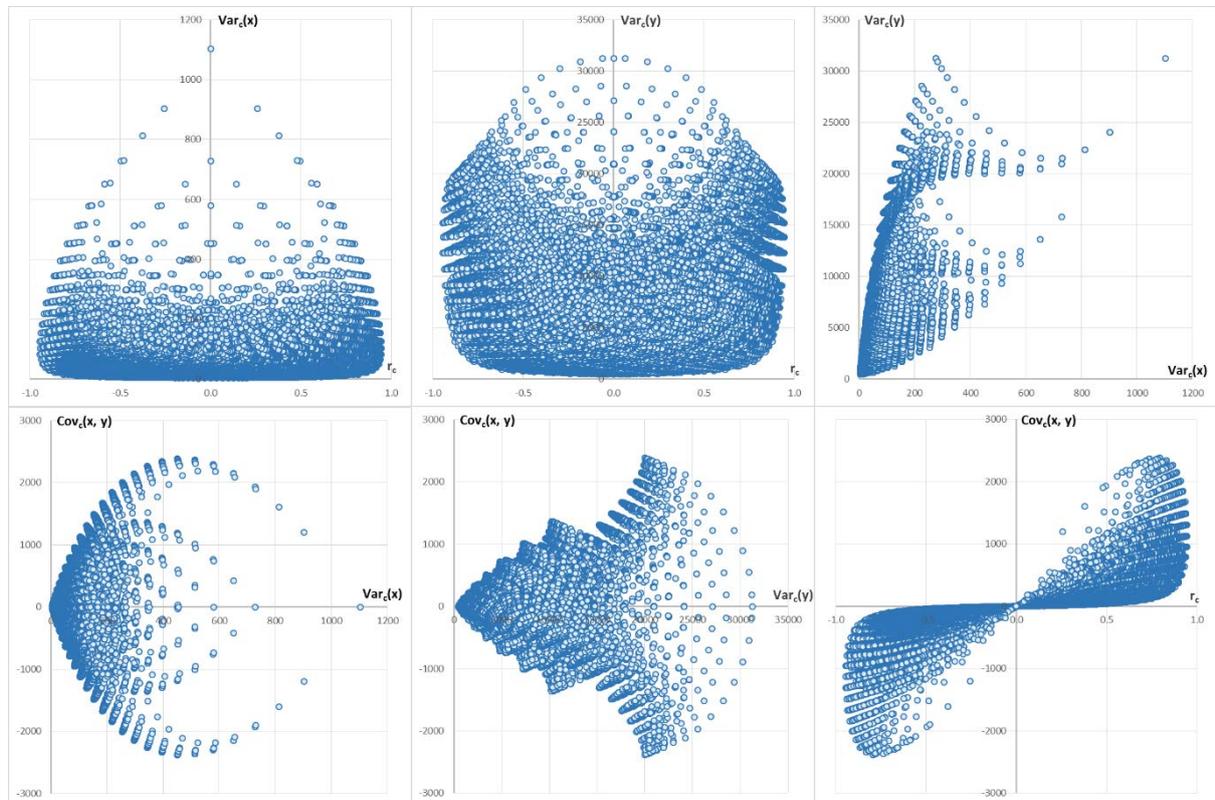

**Figure 4** The results obtained for $y = x^2$ when n = 31 and m = 2.

Figure 4 also presents the covariance clouds where the variation of compositional covariance values with compositional variances of x and y are shown. Half of the covariance values are negative and the other half are positive because of the perfect symmetry of the function. The last subfigure shows the variation of compositional covariances with the compositional correlations and the point at the origin is the Pearson's correlation value which is zero. The symmetry with respect to the origin points out the symmetry of the function $y = x^2$. Figure 5 shows the obtained results when n = 21.

As a result, these findings point out that the compositional correlation method is capable of determining the inversely and directly related portions of the examined function and provides a quantitative measure of the degree of association. For the case of $y = x^2$, the HCC and LCC values are close to the extreme values (+1 and -1) and shows that half of the function is inversely and the other half is directly related with x. The HCC and LCC values will be closer to the extreme values when n is increased. For this simple function, it is easy to see this result without making any calculations but the following examples (especially the gene expression series) show that the presented approach is also successful when the relationships cannot be determined manually.

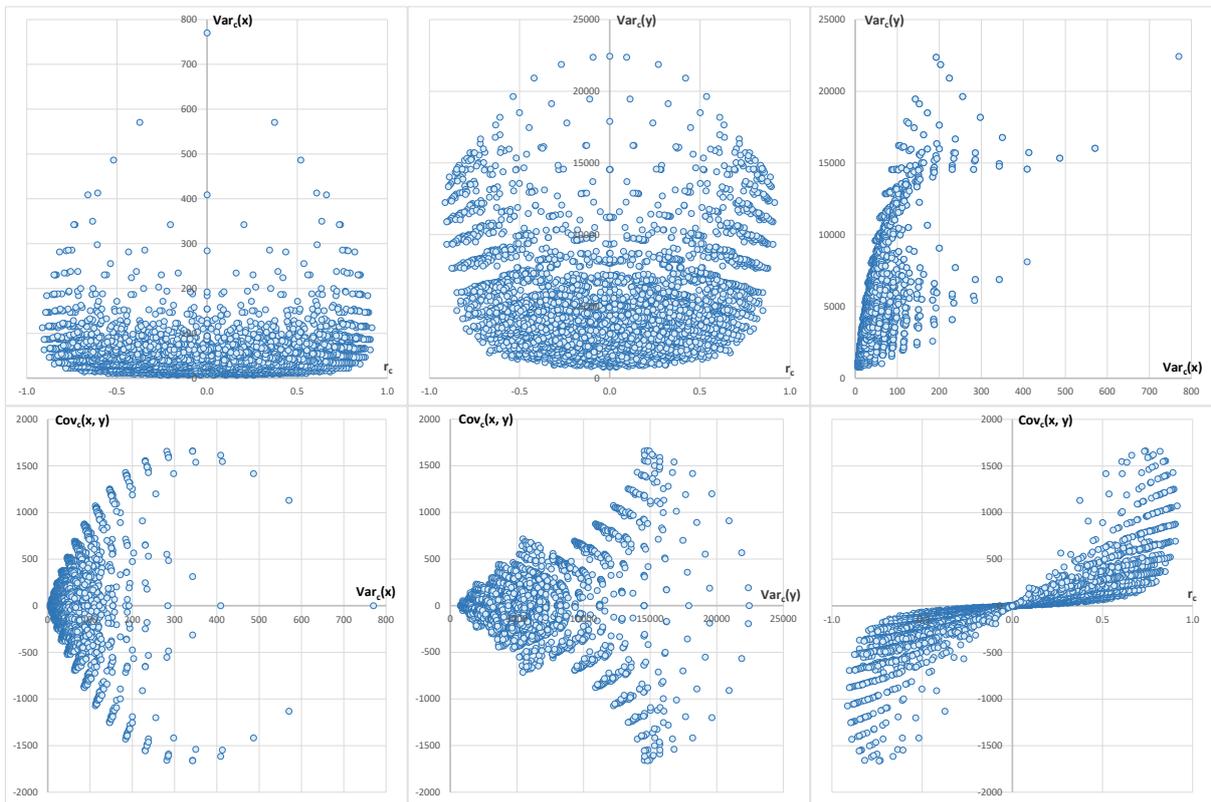

**Figure 5** The results obtained for $y = x^2$ when n = 21 and m = 2.

### 2.3.2 Results for $y = x^3 - x$

Figure 6 summarizes the results obtained for $y = x^3 - x$. For this function the majority of the compositional correlations have a positive value because the length of the increasing curve is higher than the length of the decreasing curve (the directly related portion is longer than the inversely related portion as shown in Figure 3). The variance and covariance clouds are not symmetric across the vertical axis. The HCC value (0.9397) is obtained for the composition [8, 2, 2, 2, 2, 2, 2, 2, 9] (the BCC) and the LCC value (-0.8341) is obtained for the composition [2, 2, 2, 2, 15, 2, 2, 2, 2] (the WCC). The determined BCC and WCC compositions exactly indicate the directly and inversely related regions of the function. Most of the covariance cloud is in the positive region. The Pearson's correlation between the function y and x is 0.4210 and does not provide any information about the structure of association.

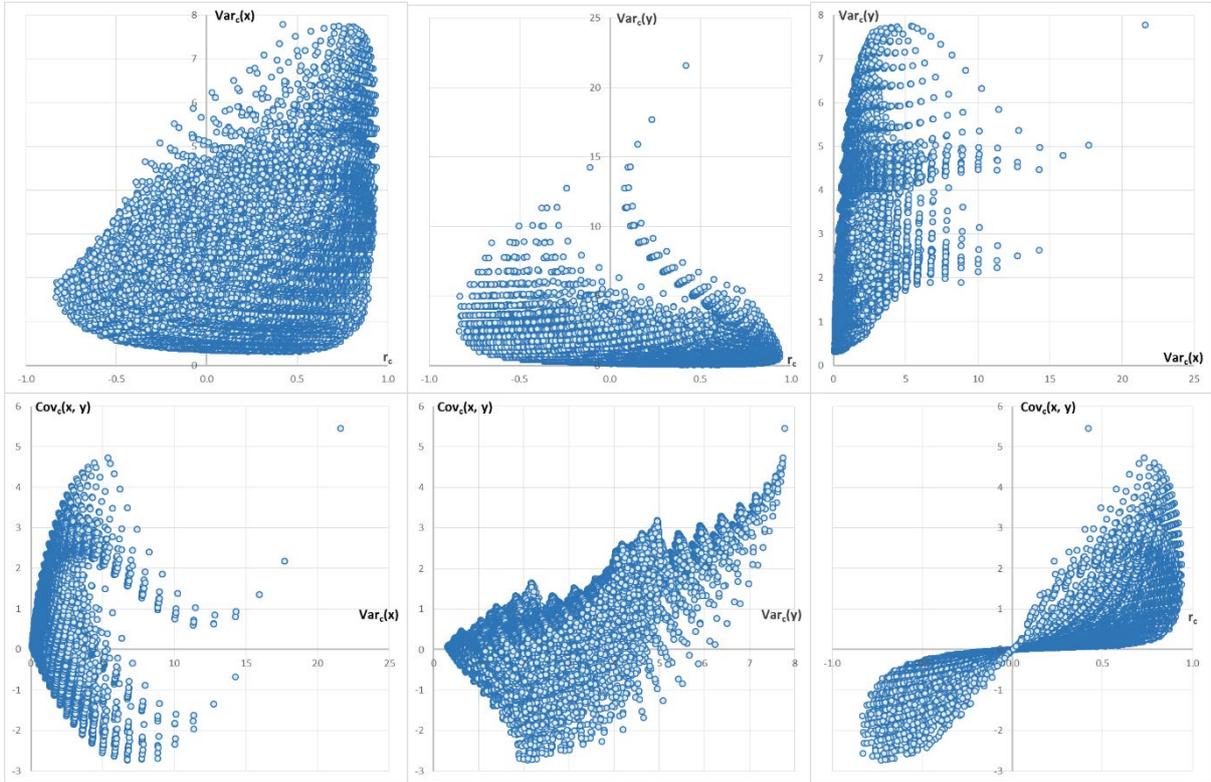

**Figure 6** The results obtained for y = x³ – x when n = 31 and m = 2.

### 2.3.3   Results for y = x⁴ – 10x² + 9

The results obtained for the function $x^4 - 10x^2 + 9$ are summarized in Figure 7. The variance and covariance clouds are symmetric as expected. The Pearson's correlation for this function is also zero as was the case for $y = x^2$ and it is very far from representing the relationship between y and x (a zero correlation is claimed to be an indicator of non-associatedness). For this function, the HCC (0.7944) is obtained for the composition [2, 2, 14, 2, 2, 2, 5, 2] (the BCC) and the LCC (-0.7944) is obtained for the composition [2, 5, 2, 2, 2, 14, 2, 2]. The compositional correlations for this function is in a narrower range then it was for the function $y = x^2$.

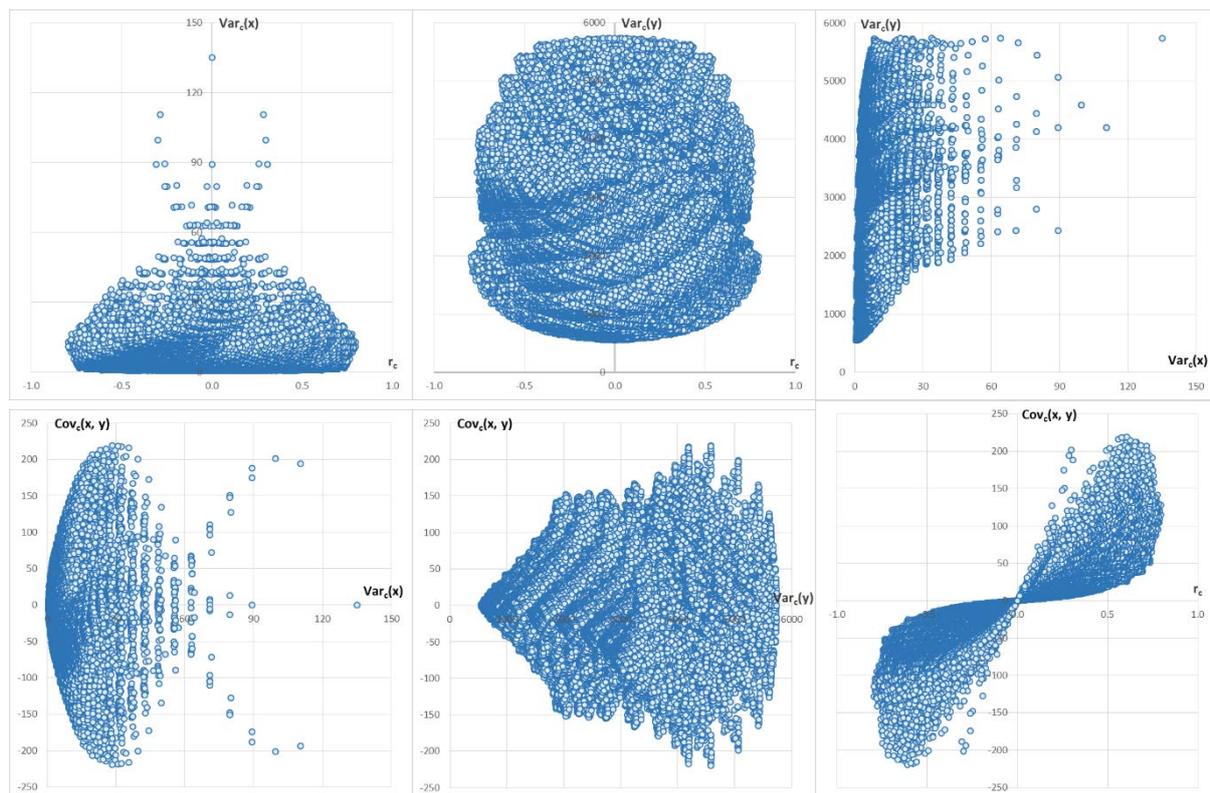

**Figure 7** The results obtained for y = x⁴ – 10x² + 9 when n = 31 and m = 2.

### 2.3.4    Results for y = x³ + x² + 2x + 4

The curve of y = x³ + x² + 2x + 4 is closer to linearity when compared with the above three polynomial functions. The results shown in Figure 8 also validate this situation. The compositional covariance clouds generated by comparing with the compositional variances do not spread out from linearity as it was the case for the above functions. In fact, all the compositional covariance values lie on a straight line when both the compared functions are linear and all compositional correlation values become equal to 1 (the compositional variances also lie on a straight line in this case). It must also be noted that, for this function, all compositional correlation values range between 0.9753 and 0.6107 showing that all compositions of the y function are in direct relationship with x and that the relationship is strong all through the investigated range of the function.

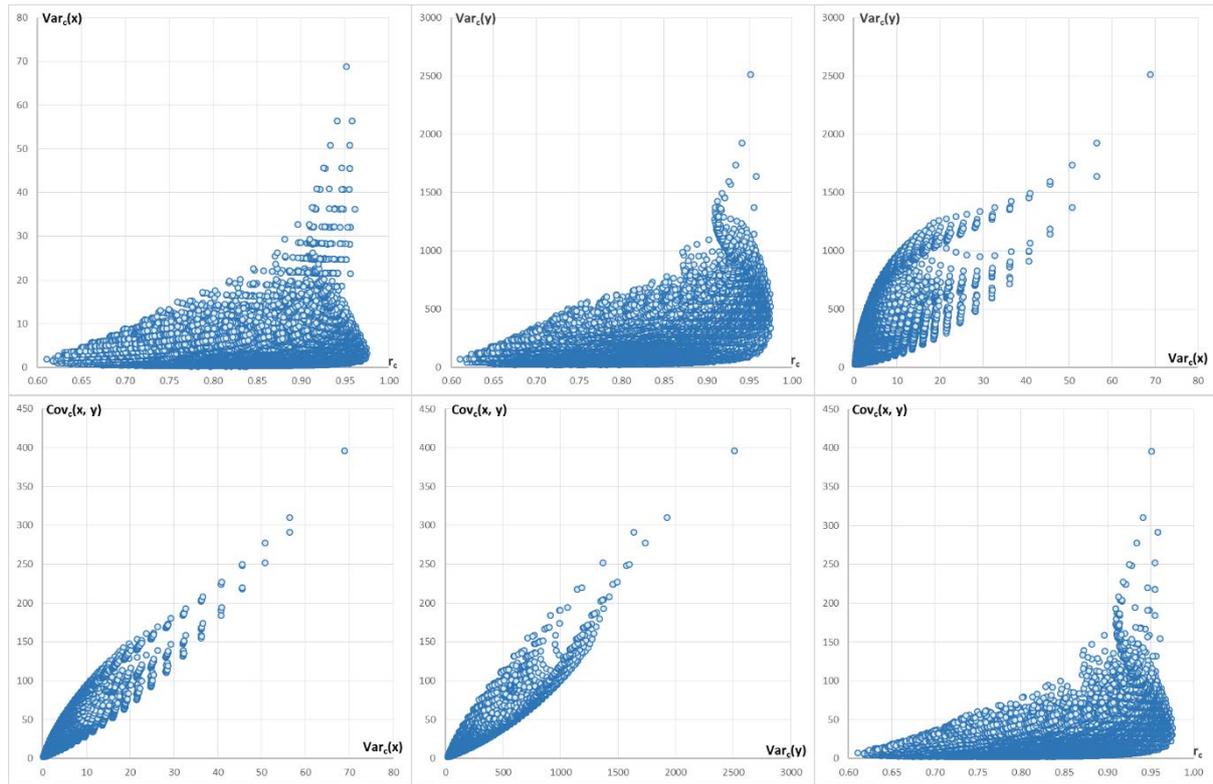

**Figure 8** The results obtained for y = x³ + x² + 2x + 4 when n = 31 and m = 2.

## 2.4 The CompCorr Software

A software named CompCorr is developed in Python for implementing the introduced compositional correlation method. The software accepts an Excel file containing the data series as input and generates a text file as output containing the compositional correlations. The number of compositional correlations calculated for each pair varies according to the length of the data series and the minimum number of accepted values in each part of the compositions. The compositions were determined by using the ruleGen function which generates all interpart restricted compositions of n by using restriction function sigma (Kelleher 2005). For each composition, the compositional correlation is determined by using Equation 1. The CompCorr code (Online Resource 2) provided together with this manuscript calculates all compositional correlations between the genes YLL024C and YLL026W and writes them to the file Output.Spellman.YLL024C.YLL026W.n23.m2.txt.

## 3. RESULTS

### 3.1 Application on gene expression data

Gene expression over time is a continuous process and can be considered as a continuous curve or function (Coffey and Hinde 2011). Genome-wide association studies try to determine associated gene pairs by comparing the expression series of each gene (Zhang et al., 2009). Understanding the temporal relationships between the expression profiles of genes is crucial in:

— determining the causes, functions and consequences of the biological processes like the cell cycle (Dancik and Theodorescu 2014; Martinez et al. 2015; Rustici et al. 2004; Sherr 1996; Spellman et al. 1998; Tusher et al. 2001),

— identifying the roles of genes in the stages of developmental processes of organisms Bird 2002 (Bird 2002; Both et al. 2005; Brand and Perrimon 1993; Brunskill et al. 2014; Li et al. 2015; Logan and Nusse 2004; Scheffer et al. 2015; Song et al. 2015),

— investigating the functions of individual genes by exploring genetic interactions and applying knockouts (Belle et al. 2015; Chen et al. 2004; González et al. 2014; Han et al. 2004; Kesarwani et al. 2007; Yan et al. 2001; Zalc et al. 2015),

— developing drugs to cure diseases by identifying genes that act in response to a certain disease (Bar-Joseph 2004; Caputo et al. 2008; César-Razquin et al. 2015; Comer et al. 2015; Ebert et al. 2009; Giaccia et al. 2003; Hanash 2003; Perros 2015; Zhang et al. 2010; Zon and Peterson 2005).

The developed method was used for calculating all compositional correlations between all possible pairs of 4381 budding yeast (saccharomyces cerevisiae) genes. Compositional correlations against time were also calculated for each gene. The evaluated gene expression dataset consists of the results of the cdc15 experiment made by Spellmann et al. (1998). The expression data provided by Reshef et al. (2011) was used in this study. The expression time series for each gene consists of 23 observations (n = 23) and the number of all possible gene pairs is 9594390. For each gene pair, all possible compositional correlations were calculated by considering the minimum number of observations in each part of a composition to be at least 4 (a total of 250 compositional correlations for each pair when m = 4). The total number of calculated compositional correlations is over 2.3 billion. All of the calculated compositional correlations were not stored as output because this would significantly slow down the file generation process and increase the requirement of storage space. Only HCC, r, LCC, BCC and WCC values for all possible data series pairs were written to the output file. The output file containing the HCC, r, LCC, BCC and WCC for each gene pair is provided for download via the following link for enabling further investigation for the gene researchers. The file contains a huge amount of clues for determining the relationships and functions of the hundreds of yeast genes which are still unidentified.

https://www.dropbox.com/s/lqqnmafx9h6g1rr/Compositional.Correlations.Spellman. m4.rar?dl=0

Among all the compared gene pairs, the highest compositional correlation (0.9928) was obtained between the genes YDL003W and YDR097C for the composition [7, 4, 8, 4] (Table S1 in Online Resource 1). For this gene pair, r = 0.9851 and LCC = 0.9422 and the small difference between HCC and LCC indicates a very strong relationship all through the observed period (Figure S1 in Online Resource 1). LCC is higher than 0.9 for 777 gene pairs while it is over 0.85 for 5202 gene pairs.

The lowest compositional correlation (-0.9822) was obtained for the pair YIL141W and YMR031C for the composition [9, 4, 5, 5] (Table S2 in Online Resource 1). For this pair, r = -0.9319 and HCC = -0.7912 (Figure S2 in Online Resource 1). HCC is lower than -0.9 for 146 gene pairs while it is less than -0.85 for 1355 gene pairs. The HCC values are over 0.9 for 31185 (0.325%) gene pairs (Table S1 in Online Resource 1) while the Pearson's correlation is over 0.9 for only 2684 (0.0027%) of the gene pairs. For example,

HCC = 0.93 for the composition [5, 4, 7, 7] of the gene pair YHR145C and YIL093C while the composition [23] is the WCC and gives LCC = 0.00 which is the Pearson's correlation (Figure 9).

This gene pair is among the 58 gene pairs for which the HCC > 0.9 while -0.1 < r < 0.1 (Table S3 in Online Resource 1). Figures 9b - 9e show the expressions of the genes for each part of the BCC and Figure 9f shows the expression values together with the Pearson's correlations for each part of the BCC. Even though the Pearson's correlation for the whole series is zero, the Pearson's correlations for each part of the BCC are very high (between 076 and 0.96; 3/4 being over 0.9). This clearly shows that the Pearson's correlation for the gene pair is misleading because there seems to be a very strong direct relationship between the genes as shown by the time series graph and the very high compositional correlation value.

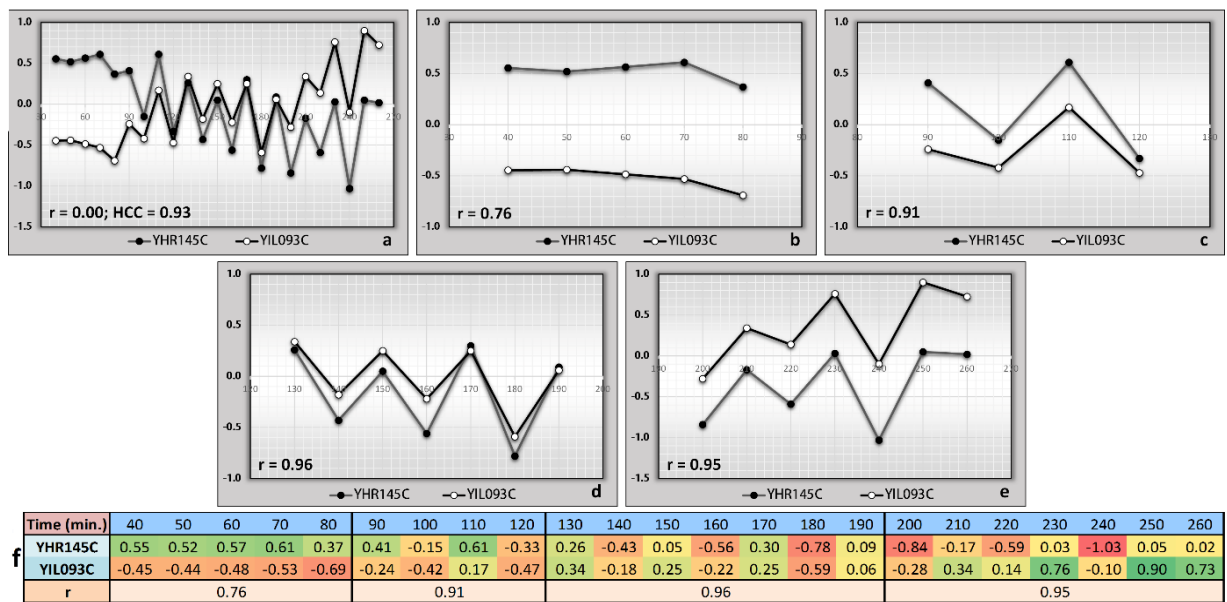

| Time (min.) | 40 | 50 | 60 | 70 | 80 | 90 | 100 | 110 | 120 | 130 | 140 | 150 | 160 | 170 | 180 | 190 | 200 | 210 | 220 | 230 | 240 | 250 | 260 |
|---|---|---|---|---|---|---|---|---|---|---|---|---|---|---|---|---|---|---|---|---|---|---|---|
| YHR145C | 0.55 | 0.52 | 0.57 | 0.61 | 0.37 | 0.41 | -0.15 | 0.61 | -0.33 | 0.26 | -0.43 | 0.05 | -0.56 | 0.30 | -0.78 | 0.09 | -0.84 | -0.17 | -0.59 | 0.03 | -1.03 | 0.05 | 0.02 |
| YIL093C | -0.45 | -0.44 | -0.48 | -0.53 | -0.69 | -0.24 | -0.42 | 0.17 | -0.47 | 0.34 | -0.18 | 0.25 | -0.22 | 0.25 | -0.59 | 0.06 | -0.28 | 0.34 | 0.14 | 0.76 | -0.10 | 0.90 | 0.73 |
| r | | | | | 0.76 | | | | 0.91 | | | | | | | 0.96 | | | | | | | 0.95 |

**Figure 9 (a)** Expression time series for the genes YHR145C and YIL093C; **(b)** first 5 expression pairs (40 to 80 minutes); **(c)** 4 expression pairs from 90 to 120 minutes; **(d)** 7 expression pairs from 130 to 190 minutes and **(e)** 7 expression pairs from 200 to 260 minutes and **(f)** the expression series and the correlations between the parts of the BCC of the genes YHR145C and YIL093C.

After determining the strong direct relationship between the genes YHR145C and YIL093C by considering m = 4 in the calculation of all the possible 250 compositional correlations (Table S4 in Online Resource 1), the 1278 compositional correlations for m = 3 (Table S5 in Online Resource 1) and 17711 compositional correlations for m = 2 (Table S6 in Online Resource 1) are calculated separately for investigating the variation of compositional correlation for the gene pair. As the compositions for m = 3 and m = 4 are subsets of the compositions for m = 2, their compositional correlation values remain within the compositional correlation range obtained for m = 2 (Figures 10a, 10b and 10c).

All compositional correlations are higher than r (which is 0.00) when m = 3 and m = 4. Similarly, when m = 2, 17709 of the 17711 compositional correlations (99.99%) are higher than r (3773 of them (21.3%) are over 0.9, 11895 of them (67.2%) are over 0.8 and 16845 of them (95.1%) are over 0.6). The results for m = 2 and m = 3 show that the obtained results for m = 4 provide sufficient information on the compositional correlation structure of the investigated gene expression series. This indicates that there is a strong direct numerical relationship between the expressions which might point out a functional relationship even though the Pearson's correlation is zero.

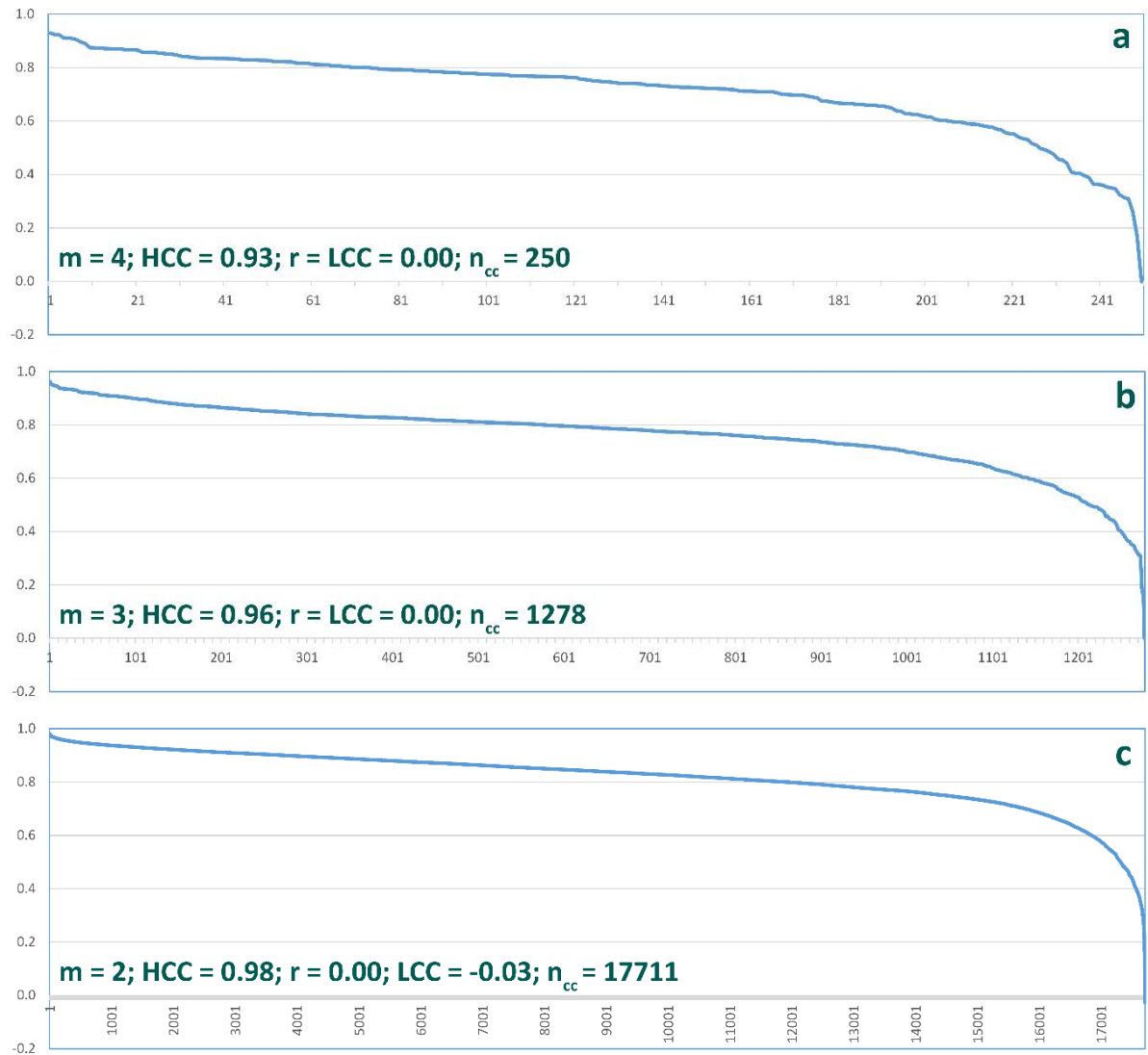

**Figure 10** The compositional correlations obtained for the gene pair YHR145C and YIL093C when m = 4 **(a)**, m = 3 **(b)** and m = 2 **(c)**. The HCC, r, LCC and the number of compositional correlations ($n_{cc}$) are indicated on each figure.

The variance and covariance clouds of the genes YHR145C and YIL093C shown in Figure 11 also validate that the Pearson's correlation is far from representing the association between the genes. The Pearson's correlation is a point at the far end (the point at the origin) of the tail of the compositional covariance cloud shown in the bottom right panel and it is far from representing the strong direct (positive) relationship between the genes shown by all the panels in the figure.

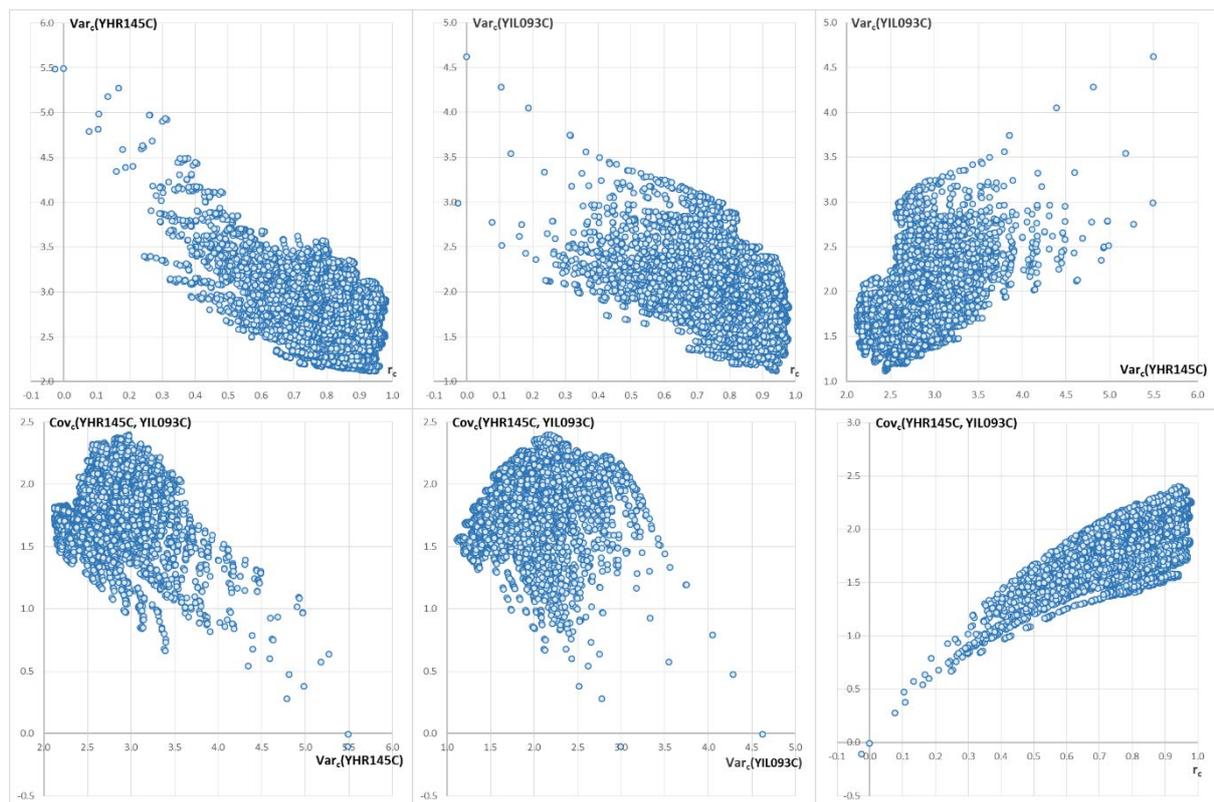

**Figure 11** The results obtained for the gene pair YHR145C and YIL093C when n = 23 and m = 2.

## 3.2 Inverse relationships

As in all association studies, determination of inverse relationships between genes might also be as important as determining direct relationships for assessing expression balance of proteins (Subbarayan et al. 2005). Pearson's correlation is below -0.9 for 554 of the investigated gene pairs, while 12373 gene pairs have a LCC value below -0.9 indicating that there might be much more inversely related yeast gene pairs than the Pearson's correlation points out (Table S2 in Online Resource 1).

The genes YBR146W and YJR045C are one of the many inversely related gene pairs that Pearson's correlation fails to detect. For this pair, HCC = r = 0.00 while LCC = -0.93 for the WCC which is [4, 4, 6, 4, 5]. Figure 12 shows the expression time series for this pair together with the correlations for each part of the WCC. The negative correlations for each part are very close to -1 indicating a strong inverse relationship but the combined parts produces a zero Pearson's correlation falsely proposing that the genes have no relation. In fact, this inverse relationship is clearly shown by the negative Pearson's correlation values (ranging between -0.88 and -1.00) calculated for each part of the WCC. The inverse relationship is also apparent in the time series graph but it is practically impossible to generate graphs and determine these types of relationships manually when there are millions of pairs of genes. The computational procedure presented in this paper enables determination of these relationships by calculating compositional correlations for all possible compositions.

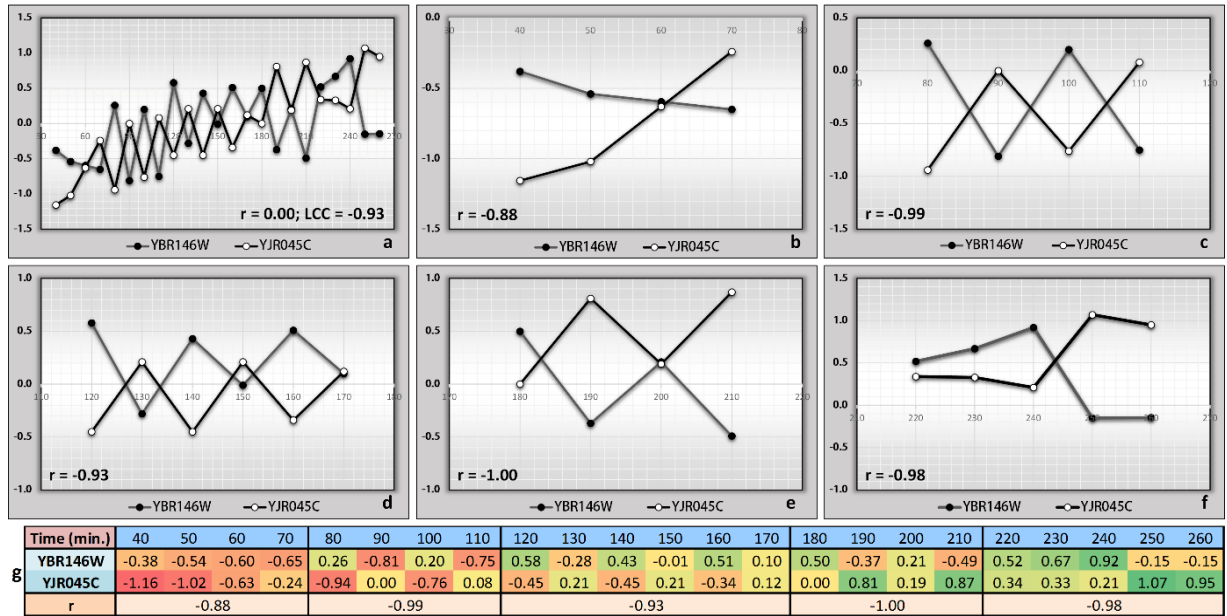

**Figure 12 (a)** Expression time series for the genes YBR146W and YJR045C; **(b)** first 4 expression pairs (40 to 70 minutes); **(c)** 4 expression pairs from 80 to 110 minutes; **(d)** 6 expression pairs from 120 to 170 minutes; **(e)** 4 expression pairs from 180 to 210 minutes and **(f)** 5 expression pairs from 220 to 260 minutes and **(g)** the expression series and the correlations between the parts of the WCC of the genes YBR146W and YJR045C.

Tables S7, S8 and S9 (Online Resource 1) respectively show in ascending order, the compositional correlations obtained for the genes YBR146W and YJR045C by taking m = 4, m = 3 and m = 2. When m = 3 and m = 4, all compositional correlations are lower than r (which is 0.00) and when m = 2, 17709 of the 17711 compositional correlations (99.99%) are lower than r (3783 of them (21.4%) are under -0.9, 12315 of them (69.6%) are under -0.8 and 17130 of them (96.7%) are under -0.6) (Figure 13a, 13b and 13c). These results imply that these genes might have a strong inverse functional relationship even though Pearson's correlation is zero.

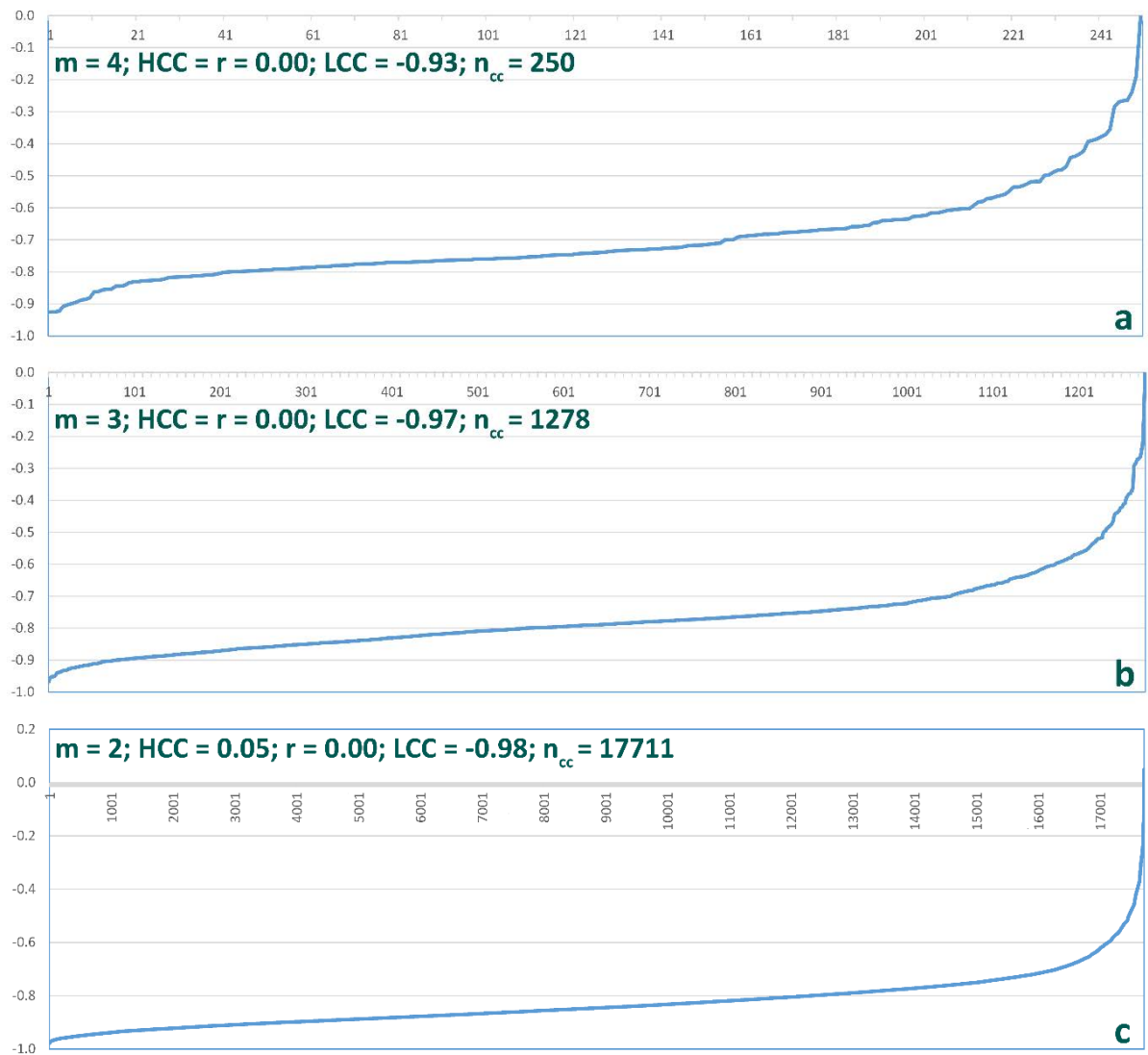

**Figure 13** The compositional correlations obtained for the gene pair YBR146W and YJR045C when m = 4 (a), m = 3 (b) and m = 2 (c). The HCC, r, LCC and the number of compositional correlations ($n_{cc}$) are indicated on each figure

The variance and covariance clouds of the genes YBR146W and YJR045C shown in Figure 14 show that the Pearson's correlation does not correctly point out the association between the genes. The Pearson's correlation is a point at the far end (the point at the origin) of the tail of the compositional covariance cloud shown in the bottom right panel and it is far from representing the strong direct (positive) relationship between the genes shown by all the panels in the figure.

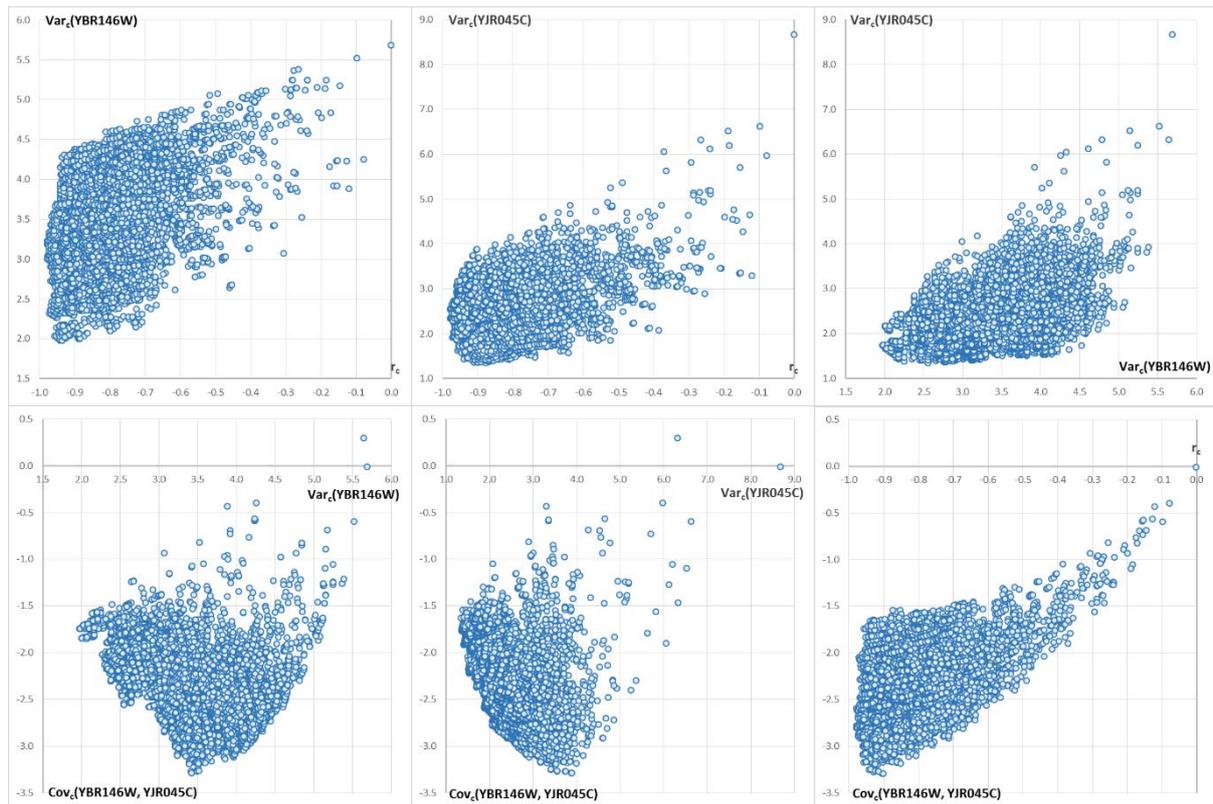

**Figure 14** The results obtained for the genes YBR146W and YJR045C when n = 23 and m = 2.

The above gene pairs are only two examples among the thousands of pairs with probable functional relationships determined by the compositional correlation method but not by r. Some other selected examples with high compositional correlation but significantly lower r values are presented in Figure S3 (Online Resource 1). Each graph in the figure includes HCC, r and LCC values together with the BCC's and WCC's. Table S10 (Online Resource 1) shows the expression values of the gene pairs and the correlations for all parts of the BCC's for each pair in Figure S3 (Online Resource 1).

## 3.3 Directly or Inversely Related?

Another feature of compositional correlation is its ability to give clues about existence of inverse relationships when the series have a general direct relationship (or vice versa). An example of this feature is the gene pair YLL024C and YLL026W. As shown in Figure 15a, the genes in this pair both have a similarly increasing expression trend and the value of Pearson's correlation (0.86) also validates this behavior. But, the expression values are always inversely related between all observation points (when one is increasing, the other is decreasing) only except for the minutes between 180 and 190 where they both increase and between the minutes 250 and 260 where they both decrease. This inverse expression behavior is determined by looking at the differences between HCC and LCC. For this pair, the LCC value is -0.35 indicating that there might be an inverse relationship between the genes. Calculation of correlations for the parts of the WCC ([7, 4, 4, 4, 4]) of the pair also points out the inverse relationship between the genes as shown in Figures 15b-15f. The r values of the parts of the WCC varies between -0.94 and 0.28 (4/5 being negative). In fact, 10585 of the 17711 compositions (59.8%) have a negative compositional correlation when m = 2 as shown in Figure 16. The dispersion of the variance and the covariance clouds are very limited for these genes showing that both genes are closely related and the variation of the compositional correlation between 0.859 and -0.871 shows that the genes have both a positive and a negative relationship. The compositions containing long parts produce positive compositional correlations as expected (the genes have an overall similar pattern) and the compositions composed of shorter parts produce negative compositional correlations as the smaller parts are inversely related.

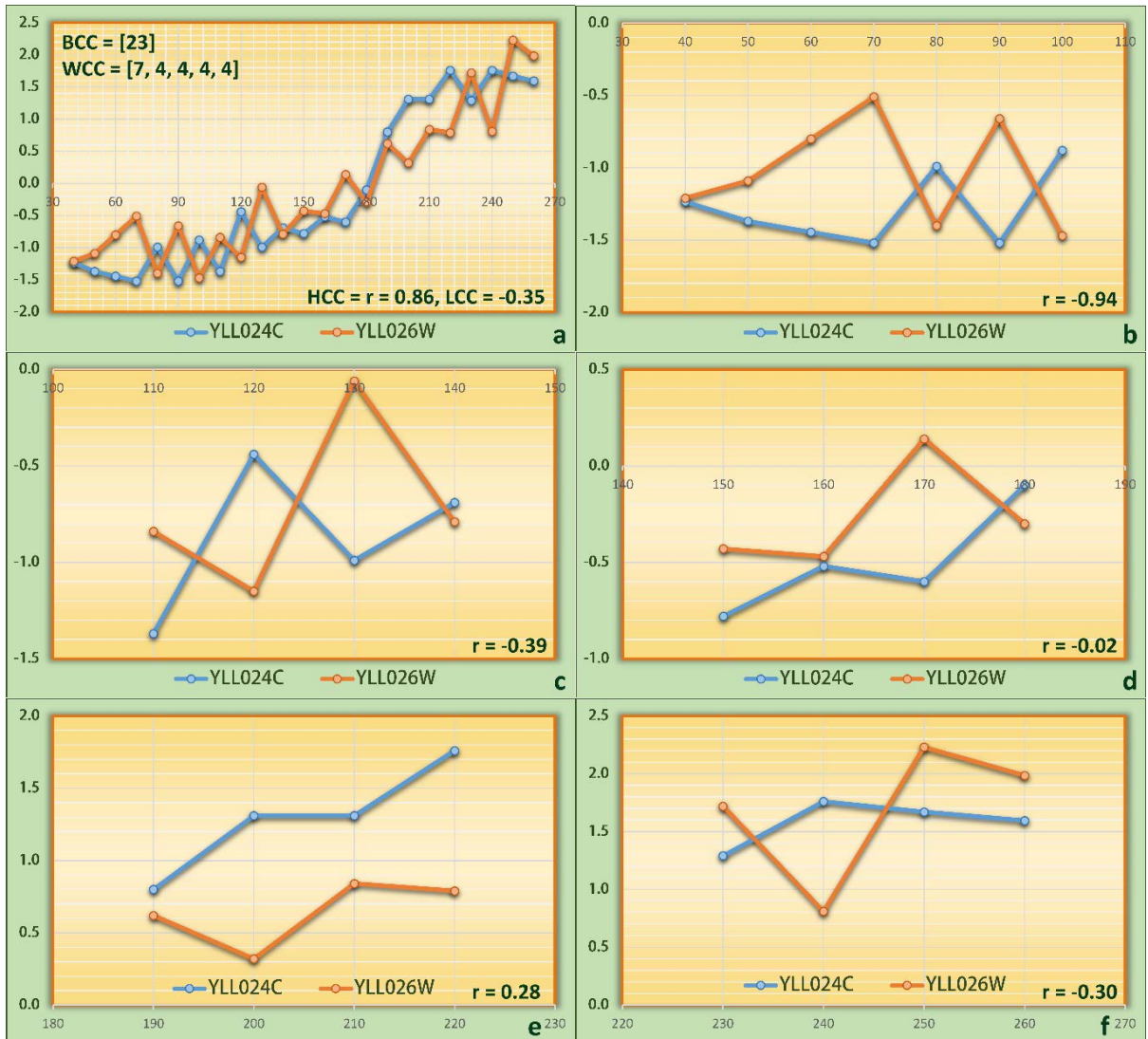

**Figure 15** The genes YLL024C and YLL026W have a similar expression trend with r=0.86 (a) but their expression patterns have inverse relationship (b-f).

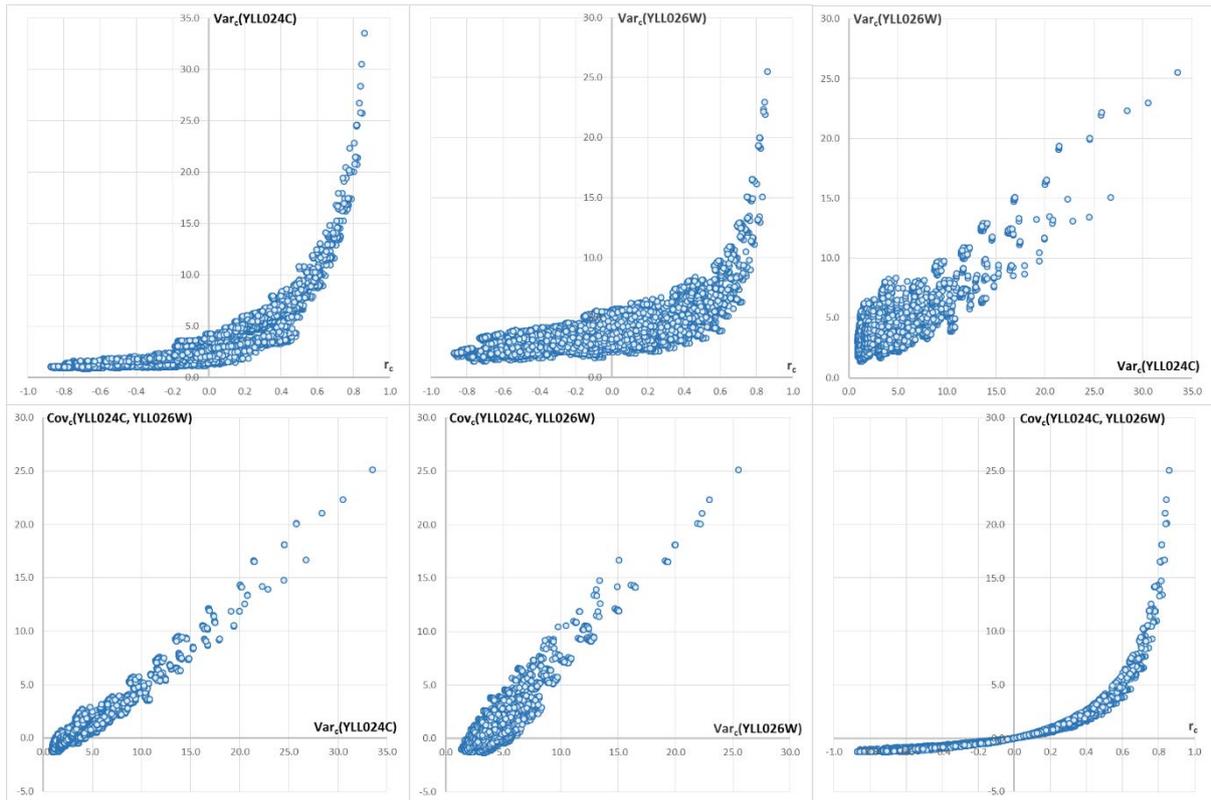

**Figure 16** The results obtained for the genes YLL024C and YLL026W when n = 23 and m = 2.

Some other selected examples of gene pairs with apparent inverse relationships which cannot be detected by calculating r are shown in Figure S4 (Online Resource 1) and the expression values of the sample pairs and the r values for each part of the BCC for each pair are presented in Table S11 (Online Resource 1). The r values of the parts (the majority being between -0.95 and -1.00) validate that the pairs might have a strong functional inverse relationship even though the r values obtained for the whole series range between -0.01 and -0.53.

### 3.4 Correlations with Time

Time always goes forward and irrespective of the differences between observation periods, it is an increasing variable generally placed on the positive x direction. Thus, the correlation between a monotonically increasing variable and time is generally a high positive number less than or equal to 1. Calculation of correlations of gene expression data with time enables detection of genes with increasing or decreasing trend but this becomes only valuable when the expression is always decreasing or increasing which is not the case for many genes. Most genes show both increasing and decreasing expression profiles through the observed course. High differences between HCC and LCC values of a gene, when compositional correlations with time are calculated, provides evidence for the existence of both increasing and decreasing periods in the expression. YJR004C (Figure 17a) is an example for such behavior which is among the genes with the highest HCC values against time. HCC for the gene is 0.93 for the composition [2, 2, 7, 2, 2, 8] and LCC is -0.84 for the composition [5, 2, 2, 7, 2, 2, 3].

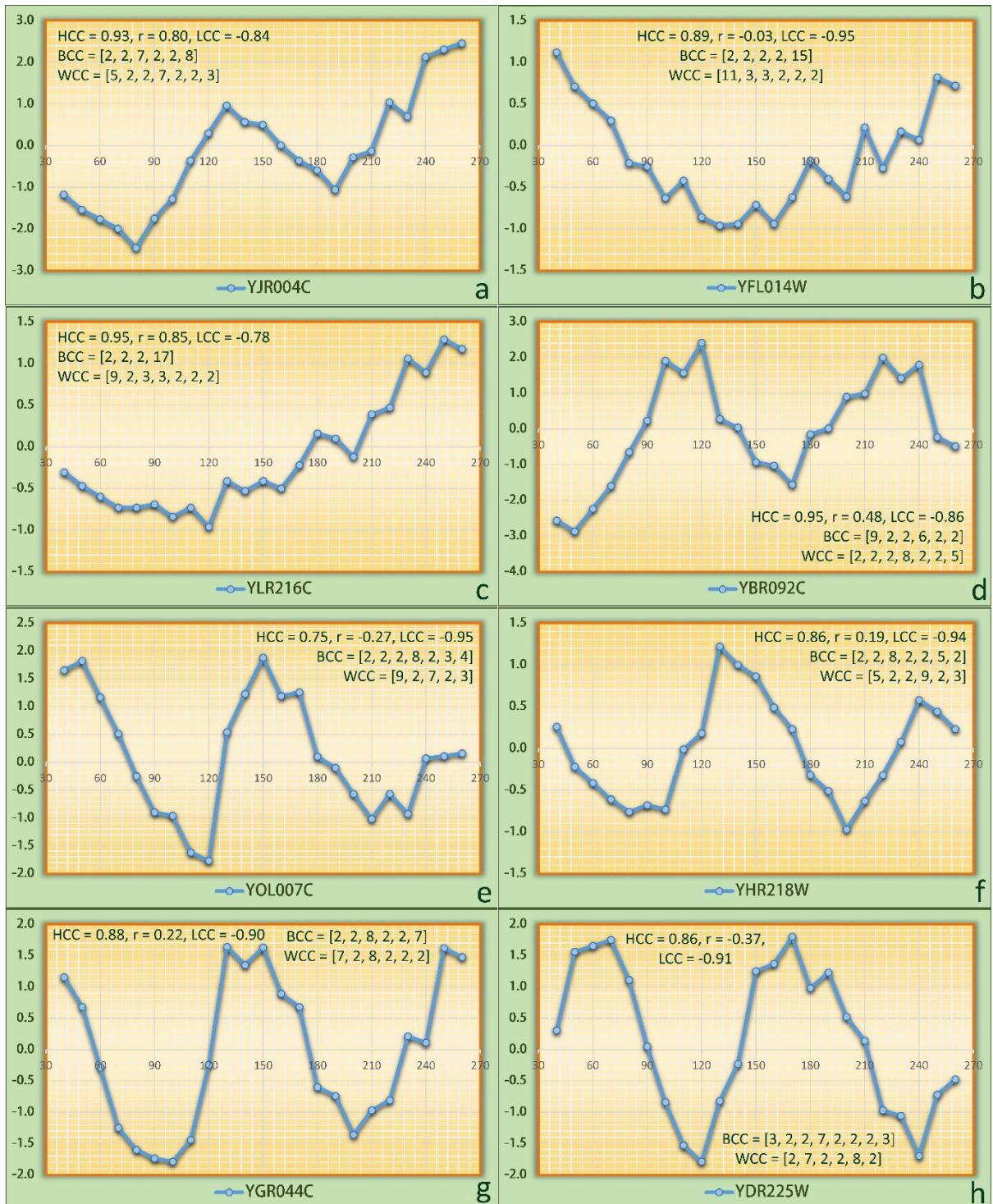

**Figure 17** Some selected genes detected by calculating compositional correlations with time. The genes have both decreasing and increasing almost linear (a, d-h) and nonlinear (b, c) parts. The HCC, r and LCC values and the best and worst correlated compositions (BCC & WCC) are also indicated on each figure.

The full list of obtained HCC, r and LCC values against time for all genes are presented in Table S12 (Online Resource 1). High differences between HCC and LCC values also enable determination of genes with nonlinear (YFL014W and YLR216C) or fluctuating (YBR092C, YOL007C, YHR218W, YGR044C, YDR225W) expression profiles having strongly linear parts (Figure 17b-17h). Among all the genes, the minimum HCC value is 0.34 for YDR199W and the highest LCC is -0.32 for YNL007C (making all HCC values positive and all LCC values negative) showing that no gene in the evaluated set has an always increasing or an always decreasing expression in the investigated period.

## 4. COMPARISONS WITH OTHER CORRELATION METHODS

Determination of genes with similar or concordant (and also inversely related) expression profiles is crucial in the determination of the functions of the genes. If a method fails to find some of the harmonious gene pairs then it will be much more difficult to make decisions on the functions of the genes as in the case of saccharomyces cerevisiae for which there are still hundreds of unidentified genes even though its genome was sequenced nearly 20 years ago (Peña-Castillo and Hughes 2007; VanderSluis et al. 2014). In this section, the performance of the compositional correlation method is compared with four widely used correlation methods which are the Pearson's correlation, Spearman's correlation, distance correlation and SGMIC which was proposed as a novel algorithm for the precise calculation of the maximal information coefficient (Zhang et al. 2014).

Genes generally exhibit varying expression behaviors through time and detecting this behavior by using standard correlation approaches is generally not possible as the series are considered as a whole and the different behaviors in subsections are not detected and considered. For example, the expressions of the gene pair YMR296C and YOL032W (Figure S5ac in Online Resource 1) first slightly decrease together between minutes 40 and 70. Then they begin to fluctuate together with a very similar pattern until minute 170. Finally, after minute 180, YOL032W shows an increasing and YMR296C shows a decreasing trend while they still have a very similar expression profile. Even though the time series graph for this pair clearly shows that there might be a strong relationship between these two genes, this behavior was not detected by Pearson's (0.048), Spearman's (0.082) and distance correlations (0.386) which pointed out a weak or no relation. The SGMIC value for this gene pair is 0.566 which seems to be a better estimate but might easily be ignored among the millions of SGMIC values for all the possible gene pairs as it also does not point out a significant relationship.

The developed compositional correlation method successfully determined the relationship between these two genes with a HCC value of 0.943 together with the 58 gene pairs with a HCC value over 0.9 for which the Pearson's correlation remains between -0.1 and 0.1 (Figure S5 in Online Resource 1). The HCC, Pearson's correlation, LCC, Spearman's correlation, distance correlation and SGMIC values for these 58 gene pairs are presented in Table 2. The minimum and maximum values for each statistic are indicated in bold.

The results show that the compositional correlations are much higher than the correlations of the compared methods only except for the SGMIC value of the pair YBR082C and YOR262W (Figure S5av in Online Resource 1). The distance correlation value (0.505) for this pair is the second highest distance correlation among the 58 pairs. It is known that the distance correlation is zero if and only if the random variables are statistically independent but it cannot be claimed that the distance correlation always exactly determines the strength of statistical dependence. The distance correlations for none of the compared pairs are close to zero showing that all of the 58 gene pairs are statistically dependent as also shown in Figure S5 (Online Resource 1).

The Pearson's and Spearman's correlations fail to detect the statistical dependence between the genes and they produce values close to zero for all the compared pairs. This result is caused by the fact that both methods generally fail to detect the relationship when the trend line for one series is increasing while the trend line for the other series is decreasing with nearly the same angle even though the series have a strong relationship. The presented gene pairs are good examples of this deficiency of Pearson's and Spearman's correlation approaches.

The best performance after the compositional correlation is shown by the SGMIC method with an average SGMIC value of 0.511 but all SGMIC values except for 0.932 obtained for the pair YBR082C and YOR262W are under 0.8 showing that the SGMIC results for these gene pairs are not sufficiently maximal to be noticed.

The presented comparative results provide sufficient proof for the statistical dependence between the compared gene pairs. There are many more gene pairs detected by the compositional correlation method with very low Pearson's correlation values close to zero. For example the number of gene pairs with a compositional correlation value over 0.8 but Pearson's correlation between -0.2 and 0.2 is 5999. Consequently, the results of this study provides the yeast researchers a very narrowed down target for defining the functions of the genes, a task which seems to be impossible by using conventional correlation measures which fail to detect relationships between genes which show alternating but dependent behavior through the course of time.

**Table 2** Comparisons of correlation methods for gene pairs with compositional correlations over 0.9 while -0.1 < r < 0.1

| Gene 1 | Gene 2 | HCC | Pearson | LCC | Spearma | Dist.Cor. | SGMIC |
|---|---|---|---|---|---|---|---|
| YCL063 | YPL203 | 0.926 | **0.098** | **0.098** | -0.031 | 0.357 | 0.546 |
| YGR244 | YKR072 | 0.908 | 0.095 | 0.095 | -0.038 | 0.363 | 0.528 |
| YLR109 | YLR441 | 0.913 | 0.094 | 0.094 | -0.028 | 0.274 | 0.445 |
| YDR375 | YDR449 | 0.906 | 0.093 | 0.093 | 0.026 | 0.326 | 0.548 |
| YLL034 | YPL118 | 0.908 | 0.091 | 0.091 | 0.029 | 0.297 | 0.652 |
| YMR09 | YNL252 | 0.905 | 0.089 | 0.089 | 0.126 | 0.320 | 0.441 |
| YPR060 | YPR158 | 0.904 | 0.084 | 0.084 | 0.124 | 0.338 | 0.510 |
| YKL150 | YNL007 | 0.941 | 0.080 | -0.028 | 0.138 | 0.352 | 0.667 |
| YKL122 | YLR109 | 0.913 | 0.079 | 0.079 | -0.052 | 0.301 | 0.520 |
| YJR054 | YOL052 | 0.903 | 0.079 | 0.077 | 0.059 | 0.295 | 0.398 |
| YGR244 | YLR075 | 0.913 | 0.078 | 0.078 | 0.066 | 0.357 | 0.586 |
| YIL093C | YLR197 | 0.914 | 0.078 | 0.078 | 0.001 | 0.419 | 0.791 |
| YFR050 | YGL189 | 0.919 | 0.078 | 0.078 | 0.053 | 0.350 | 0.544 |
| YNL005 | YOR095 | 0.903 | 0.076 | 0.076 | 0.091 | 0.302 | 0.423 |
| YJL063C | YOL097 | 0.910 | 0.073 | 0.073 | 0.006 | 0.353 | 0.423 |
| YIL070C | YLR344 | 0.907 | 0.072 | 0.072 | -0.036 | 0.291 | 0.545 |
| YLR203 | YPL048 | 0.916 | 0.069 | 0.069 | 0.072 | 0.380 | 0.464 |
| YGL049 | YMR18 | 0.905 | 0.068 | 0.068 | 0.048 | 0.252 | 0.361 |
| YGL120 | YJL063C | 0.901 | 0.068 | 0.068 | 0.015 | 0.400 | 0.559 |
| YJL125C | YNL252 | 0.936 | 0.064 | 0.064 | 0.019 | 0.329 | 0.360 |
| YDR489 | YPR158 | 0.903 | 0.063 | 0.063 | 0.065 | 0.270 | 0.418 |
| YLR354 | YNL007 | 0.912 | 0.063 | 0.033 | 0.133 | 0.317 | 0.586 |
| YEL039 | YER027 | 0.905 | 0.058 | -0.112 | 0.192 | 0.289 | 0.455 |
| YGR244 | YOR300 | 0.902 | 0.051 | 0.051 | 0.062 | 0.309 | 0.456 |
| YGR244 | YHR145 | 0.936 | 0.050 | 0.050 | -0.027 | 0.447 | 0.735 |
| YNL135 | YOR325 | 0.921 | 0.050 | 0.050 | 0.076 | 0.308 | 0.490 |
| YER156 | YLR109 | 0.920 | 0.050 | 0.050 | -0.056 | **0.208** | 0.316 |
| YBL066 | YDR231 | 0.902 | 0.048 | 0.048 | 0.198 | 0.409 | 0.464 |
| YMR29 | YOL032 | **0.943** | 0.048 | 0.048 | 0.082 | 0.386 | 0.566 |
| YCR056 | YOL032 | **0.900** | 0.044 | 0.044 | 0.088 | 0.330 | 0.453 |
| YLR203 | YPR085 | 0.922 | 0.040 | 0.040 | 0.012 | 0.367 | 0.321 |
| YIL093C | YLR175 | 0.908 | 0.035 | 0.035 | -0.055 | 0.386 | 0.697 |
| YGL219 | YNL007 | 0.903 | 0.031 | **-0.179** | 0.132 | 0.403 | 0.436 |
| YBL101 | YLR069 | 0.923 | 0.028 | 0.028 | 0.029 | 0.366 | 0.482 |
| YLL026 | YML008 | 0.909 | 0.024 | 0.024 | -0.005 | 0.288 | 0.510 |
| YJL063C | YPR062 | 0.902 | 0.023 | 0.023 | -0.146 | 0.393 | 0.761 |
| YDL022 | YJL029C | 0.919 | 0.016 | 0.016 | -0.047 | 0.353 | 0.667 |
| YDR231 | YLR185 | 0.924 | 0.016 | 0.016 | **0.200** | 0.386 | 0.588 |
| YCR056 | YPL118 | 0.913 | 0.012 | 0.012 | -0.008 | 0.290 | 0.351 |
| YJL063C | YOR300 | 0.940 | 0.007 | 0.007 | -0.052 | 0.341 | 0.493 |
| YHR145 | YIL093C | 0.929 | -0.001 | -0.001 | -0.135 | 0.407 | 0.588 |
| YLL034 | YLR203 | **0.900** | -0.008 | -0.008 | -0.029 | 0.380 | 0.592 |
| YJL063C | YMR10 | 0.912 | -0.022 | -0.022 | -0.037 | 0.341 | 0.367 |
| YBR183 | YHR216 | 0.911 | -0.030 | -0.030 | 0.064 | **0.508** | 0.775 |
| YHL035 | YNL007 | 0.905 | -0.036 | -0.177 | 0.177 | 0.367 | 0.586 |
| YGL221 | YNL007 | 0.921 | -0.040 | -0.123 | 0.113 | 0.332 | 0.618 |
| YER049 | YGR048 | 0.901 | -0.041 | -0.041 | 0.003 | 0.309 | 0.463 |
| YBR082 | YOR262 | 0.917 | -0.044 | -0.105 | 0.044 | 0.505 | **0.932** |
| YLR109 | YPR110 | 0.925 | -0.047 | -0.047 | -0.115 | 0.241 | 0.348 |
| YCL014 | YDL110 | 0.901 | -0.060 | -0.090 | **-0.204** | 0.398 | 0.649 |
| YGR097 | YLL026 | 0.904 | -0.062 | -0.062 | -0.002 | 0.274 | 0.332 |
| YLR109 | YOR300 | 0.911 | -0.063 | -0.063 | -0.087 | 0.256 | 0.426 |
| YGR228 | YOR310 | 0.904 | -0.073 | -0.073 | -0.110 | 0.251 | **0.259** |
| YLR138 | YPL118 | 0.917 | -0.076 | -0.076 | -0.116 | 0.332 | 0.324 |
| YGR244 | YLR293 | 0.924 | -0.081 | -0.081 | -0.144 | 0.350 | 0.559 |
| YDR509 | YGR254 | 0.908 | -0.082 | -0.082 | -0.187 | 0.371 | 0.499 |
| YKL024 | YLR109 | 0.907 | -0.089 | -0.089 | -0.176 | 0.301 | 0.495 |
| YPL118 | YPR048 | 0.906 | **-0.094** | -0.094 | -0.127 | 0.293 | 0.288 |
| | Max: | 0.943 | 0.098 | 0.098 | 0.200 | 0.508 | 0.932 |
| | Min: | 0.900 | -0.094 | -0.179 | -0.204 | 0.208 | 0.259 |
| | Average | 0.913 | 0.024 | 0.010 | 0.009 | 0.340 | 0.511 |

## 5. LIMITATIONS

Requirement of a longer computation time is a limitation of the method against conventional correlation calculations where mostly a single value is calculated. Nevertheless, the required duration might be significantly decreased by increasing the minimum number of elements in each part of the compositions or by using much faster supercomputers and applying parallel programming techniques. For example, the compositional correlation calculation for the gene expression data set used in this study took 14 days on a standard desktop computer with an Intel i7-3770 processor when the number of observations in each parts was 4 (It was calculated that the computation would last 111 days on the same computer when the parts of each composition would have at least 2 observations). Even though a long computation time is required, the findings show that it is worth waiting to obtain the statistically significant relationships not found by the other faster methods.

## 6. CONCLUSION

The presented compositional correlation method is capable of determining linear, nonlinear, direct and indirect relationships between data series and has a great potential of being applied in all areas of science. In the light of the findings on various polynomial functions and the gene expression data series, it is evident that the method may enable possibilities for numerous important discoveries and will contribute to the improvement of our understanding of correlation. The author also hopes that the results presented in this manuscript will provide important clues to the yeast researchers trying to find the functions of the uncharacterized yeast genes.

## SUPPLEMENTARY MATERIAL

The supplementary pdf file containing the Figures S1 to S5, Tables S1 to S12 and the Python code of CompCorr software can be downloaded from the following link:

https://www.dropbox.com/s/ai8r590sz2e8aw6/Supplementary.Material.pdf?dl=0